\documentclass[runningheads,a4paper]{llncs}%

\usepackage{xcolor}
\usepackage[T1]{fontenc}
\usepackage{graphicx}
\usepackage{url}
\usepackage{amsfonts}
\usepackage{multicol}
\usepackage{multirow}
\usepackage[utf8]{inputenc}
\usepackage{array}
\usepackage{dirtytalk}
\usepackage{booktabs}
\usepackage{enumitem}    
\usepackage{rotating}

\usepackage[nolist]{acronym}
\expandafter\def\expandafter\ac\expandafter{\expandafter\leavevmode\ac}
\begin{acronym}
\acro{ssi}[SSI]{Self-Sovereign Identity}
\acroplural{ssi}[SSI]{Self-Sovereign Identities}
\acro{dsr}[DSR]{Design Science Research}
\acro{slr}[SLR]{Systematic Literature Review}
\acro{did}[DID]{Decentralized Identifier}
\acro{w3c}[W3C]{World Wide Web Consortium}
\acro{vc}[VC]{Verifiable Credential}
\acro{dif}[DIF]{Decentralized Identity Foundation}
\acro{zkp}[ZKP]{Zero-Knowledge Proof}
\acro{ca}[CA]{Certificate Authority}
\acroplural{ca}[CA]{Certificate Authorities}
\acro{pki}[PKI]{Public Key Infrastructure}
\acro{dpki}[DPKI]{Decentralized Public Key Infrastructure}
\acro{iiw}[IIW]{Internet Identity Workshop}
\acro{p2p}[P2P]{Peer-to-Peer}
\acro{iam}[IAM]{Identity and Access Management}
\acro{idm}[IdM]{Identity Management}
\acro{dlt}[DLT]{Distributed Ledger Technology}
\acroplural{dlt}[DLTs]{Distributed Ledger Technologies}
\acro{eu}[EU]{European Union}
\acro{ec}[EC]{European Commission}
\acro{ep}[EP]{European Parliament}
\acro{poc}[PoC]{Proof of Concept}
\acro{essif}[ESSIF]{European Self-Sovereign Identity Framework}
\acro{sdx}[SDX]{sovereign data exchange}
\acro{pet}[PET]{Privacy-Enhancing Technology}
\acroplural{pet}[PETs]{Privacy-Enhancing Technologies}
\acro{tet}[AET]{Authenticity-Enhancing Technology}
\acroplural{tet}[AETs]{Authenticity-Enhancing Technologies}

\acro{iso}[ISO]{International Organization for Standards}

\acro{dp}[DP]{Differential Privacy}
\acro{kyc}[KYC]{Know Your Customer}
\acro{aml}[AML]{Anti-Money Laundering}
\acro{gdpr}[GDPR]{General Data Protection Regulation}
\acro{dsa}[DSA]{Digital Service Act}
\acro{dma}[DMA]{Digital Market Act}
\acro{dga}[DGA]{Data Governance Act}
\acro{ffod}[FFoD]{Free Flow of Non-Personal Data Regulation}
\acro{iic}[IIC]{Industrial internet Consortium}
\acro{dssp}[DSSP]{Data Space Support Platform}
\acro{dbms}[DBMS]{Database Management System}
\acro{idsa}[IDSA]{International Data Spaces Association}
\acro{ids}[IDS]{International Data Space}
\acro{sme}[SME]{Small and Medium-Sized Enterprise}
\acro{api}[API]{Application Programming Interface}
\acro{daps}[DAPS]{Dynamic Attribute Provisioning Service}
\acro{smc}[MPC]{Secure Multi-Party Computation}
\acro{he}[HE]{Homomorphic Encryption}
\acro{ssh}[SSH]{Secure Shell}
\acro{tls}[TLS 1.2]{Transport Layer Security 1.2}
\acro{voip}[VoIP]{Voice-over-Internet-Protocol}
\acro{pgp}[PGP]{Pretty Good Privacy}
\acro{ml}[ML]{Machine Learning}
\acro{tcp}[TCP]{Transmission Control Protocol}
\acro{tor}[TOR]{The Onion Router}
\acro{ds}[DS]{Digital Signature}
\acro{iot}[IoT]{Internet of Things}
\acro{dht}[DHT]{Distributed Hash Table}
\acro{ipfs}[IPFS]{InterPlanetary File System}
\acro{pbd}[PbD]{Privacy-by-Design}
\acro{ccpa}[CCPA]{California Consumer Privacy Act}
\acro{ttp}[TTP]{Trusted Third-Party}
\acro{tee}[TEE]{Trusted Execution Environment}
\acro{fl}[FL]{Federated Learning}
\acro{se}[SE]{Searchable Encryption}
\acro{dsm}[DSM]{Digital Single Market}
\acro{cdei}[CDEI]{Center for Data Ethics and Innovation}
\acro{acapy}[ACA-Py]{Aries-Cloudagent-Python}
\acro{fss}[FSS]{Function Secret Sharing}
\acro{acc}[ACC]{Aries Cloud Controller}
\acro{acm}[ACM]{Agent Connection Manager}
\acro{aws}[AWS]{Amazon Web Services}
\acro{didcomm}[DIDComm]{DID Communication}
\acro{sse}[SSE]{Searchable Symmetric Encryption}
\acro{peks}[PEKS]{Public-Key Encryption with Keyword Search}
\acro{pp}[PP]{Privacy Policy}
\acroplural{pp}[PPs]{Privacy Policies}
\acro{von}[VON]{Verifiable Organizations Network}
\acro{sde}[SDE]{Sovereign Data Exchange}
\acro{ei}[EI]{Expert Interview}
\end{acronym}

\newcommand\sbullet[1][.5]{\vcenter{\hbox{\scalebox{#1}{\textbullet}}}}

\newcommand{\rqone}{Based on researchers and industry stakeholders, what challenges hinder \ac{sde}?}
\newcommand{\rqtwo}{Which \acp{pet} and \acp{tet} have the potential to mitigate the \ac{sde} challenges identified through RQ1?}

\newcommand{\spheadingSHORT}[2][8em]{
\rotatebox{90}{\parbox{#1}{\raggedright #2}}} 

\makeatletter
\newcommand{\interviewee}[2]{%
  \protected@edef\@currentlabel{I#1}
  \phantomsection
  Interviewee #1\label{#2}
}
\makeatother

\begin{document}
\title{Mitigating Sovereign Data Exchange Challenges: A Mapping to Apply Privacy- and Authenticity-Enhancing Technologies}%
\titlerunning{Mitigating Sovereign Data Exchange Challenges}

\author{Kaja Schmidt\inst{1}
\and Gonzalo Munilla Garrido\inst{2}
\and Alexander Mühle\inst{1}
\and Christoph Meinel\inst{1}}
\authorrunning{K. Schmidt et al.}
%
\institute{Hasso Plattner Institute, Potsdam, Germany 
\email{\{kaja.schmidt,alexander.muehle,christoph.meinel\}@hpi.de}\\ 
\and
Technical University of Munich, Munich, Germany, 
\email{\{gonzalo.munilla-garrido\}@tum.de}
}

\maketitle              
\vspace{-0.17cm}
\begin{abstract}
Harmful repercussions from sharing sensitive or personal data can hamper institutions’ willingness to engage in data exchange. Thus, institutions consider \acp{tet} and \acp{pet} to engage in \textit{\ac{sde}}, i.e., sharing data with third parties without compromising their own or their users’ data sovereignty. However, these technologies are often technically complex, which impedes their adoption. 
To support practitioners select \acp{pet} and \acp{tet} for \ac{sde} use cases and highlight \ac{sde} challenges researchers and practitioners should address, this study empirically constructs a challenge-oriented technology mapping. 
First, we compile challenges of \ac{sde} by conducting a systematic literature review and expert interviews. Second, we map \acp{pet} and \acp{tet} to the \ac{sde} challenges and identify which technologies can mitigate which challenges. We validate the mapping through investigator triangulation. 
Although the most critical challenge concerns data usage and access control, we find that the majority of \acp{pet} and \acp{tet} focus on data processing issues.


\end{abstract}

\acresetall
\section{Introduction}
\vspace{-0.1cm}

Companies seeking growth collect and analyze increasing amounts of data to innovate and improve their products and services. Thereby, data becomes a valuable resource~\cite{cappiello.etal_2020_data}. 
However, despite the upside potential of collecting more data, institutions are reluctant to share (often sensitive) information with third parties as they may lose control over who and how their data is accessed, used, or distributed~\cite{bastiaansen.etal_2020_usercentric,gil.etal_2020_evaluation}. 
European initiatives (e.g., Gaia-X~\cite{bmwi_2020_project}, \ac{idsa}~\cite{otto.etal_2017_reference}) were formed to entice the industry into considering data sharing without risking unauthorized distribution or misuse of data (i.e., data sovereignty).
We refer to \emph{\ac{sde}} as the ability of a digital subject to share their data with third parties without compromising their data sovereignty, i.e., the self-determination over accessing, processing, managing, or securing their data.

As institutions work towards \ac{sde}, \acp{pet} and \acp{tet} receive increasing attention. \acp{pet} manage or modify data to protect sensitive personal information~\cite{hes.etal_1998_privacyenhancing} and \acp{tet} enhance authenticity, and integrity of data and information in a system~\cite{munilla-garrido.etal_2021_revealing} to incentivize data sharing~\cite{schmidt.etal_2021_exploitative}. Some \acp{pet} can also be \acp{tet}, but  not all \acp{tet} are \acp{pet} and vice versa~\cite{munilla-garrido.etal_2021_revealing}.
However, researchers and practitioners struggle to understand the technologies due to their technical complexity, low maturity, the wide range of possible variations and combinations, and potential economic risks~\cite{munilla-garrido.etal_2021_revealing,zrenner.etal_2019_usage,zoll.etal_2021_privacysensitive}. While researchers focused on \ac{sde} challenges~\cite{bastiaansen.etal_2020_usercentric,lauf.etal_2022_linking,zrenner.etal_2019_usage} or capabilities of \acp{pet} and \acp{tet}~\cite{pennekamp.etal_2019_dataflow,danezis.etal_2014_privacy}, we are the first to help practitioners select \acp{pet} or \acp{tet} according to the challenges in \ac{sde} use cases.

We build our contribution by first identifying the challenges of \ac{sde} with a \ac{slr} and \acp{ei}. 
Consecutively, we map \acp{pet} and \acp{tet} against the challenges they can tackle via investigator triangulation. 
We provide four contributions. We (1) compile a list of relevant \emph{\ac{sde} challenges} (Section~\ref{sec:findings-rq1}), (2) thoroughly research \acp{pet} and \acp{tet} relevant to mitigate \ac{sde} challenges (Section~\ref{sec:background}) in a \emph{technology classification}, (3) map \acp{pet} and \acp{tet} against the \ac{sde} challenges, summarized into a \emph{technology mapping}, and (4) outline five \emph{key findings} (Section~\ref{sec:discussion}). Our study thereby synthesizes previous research in an actionable way: 
the technology mapping is a challenge-oriented approach supporting the identification of appropriate \acp{pet} and \acp{tet} to tackle \ac{sde} challenges. 
\section{Research Methods}
\label{sec:methology}

\subsection{Methodologic Triangulation}
\label{sec:method-rq1}

\emph{\textbf{RQ1:} \rqone}
We conduct an \ac{slr} and complement the findings through \acp{ei} to answer RQ1. We consolidate the results into $13$ challenges of \ac{sde} (Table~\ref{tab:rq1-slr-appendix}), classified into organizational issues, data processing and publishing, and infrastructure challenges (Table~\ref{tab:rq1-challenges}). 

\vspace{-1em}
\subsubsection{Systematic Literature Review}
\label{sec:method–slr}

The \ac{slr} process consists of three phases~\cite{vombrocke.etal_2009_reconstructing,webster.watson_2002_analyzing}, which are documented to increase transparency and reproducibility of the \ac{slr}. 
First, we define a search strategy by choosing search terms and databases. We apply the search term \emph{\say{data sovereign*}} to multiple databases (Table~\ref{tab:slr-search-strategy}) to ensure a coverage of software engineering
and information systems literature
As of April 2021, we identified $205$ search hits. 
Second, we define inclusion criteria and exclusion criteria. At first, we included all publications satisfying formal requirements (e.g., unique English journal and conference hits). 
Then, we exclude articles based on titles and abstracts (e.g., unrelated data sovereignty subject, too broad focus) followed by exclusion based on full-test analysis (e.g., no data exchange focus, generic challenges of technologies). We exclude $191$ hits, resulting in a subset of 14 final hits (6.8\% relevance rate). We identified $2$ more hits from the backward search. The final hits were published between 2014 and 2021.
Third, we analyze the final hits using content analysis. We extract, summarize, thematically compare, and theoretically generalize text passages describing \ac{sde} challenges~\cite{meuser.nagel_2009_expert}. 
Overall, 13 challenges are highlighted~(Table~\ref{tab:rq1-slr-appendix}).

\begin{table}[t]
\caption{Selection Process of the SLR}
\label{tab:slr-search-strategy}
\centering
\scriptsize
\begin{tabular}{l >{\centering\arraybackslash}p{0.19\textwidth} >{\centering\arraybackslash}p{0.19\textwidth} >{\centering\arraybackslash}p{0.17\textwidth} c}
    \toprule
    \multirow{2}{*}{\textbf{Selection Criteria}} & \textbf{ACM Digital Library}\textsuperscript{a} & \textbf{IEEE Xplore Digital Library}\textsuperscript{b} & \textbf{Web of Science}\textsuperscript{c} & \multirow{2}{*}{\textbf{Total}} \\ \midrule
    Total search hits & 71 & 49 & 85 & 205 \\
    Unique journal/conference hits & 49 & 64 & 26 & 139 \\
    Title and abstract hits & 17& 14& 15&46 \\
    Full-text hits & 5& 4& 5& 14 \\ \midrule
    Backward search final hits & -& -& -& 2\\ \midrule
    \textbf{Total final hits} & -&- &- & \textbf{16}\\
    \bottomrule
    \multicolumn{5}{l}{\textsuperscript{a}\url{https://dl.acm.org/}, \textsuperscript{b}\url{https://ieeexplore.ieee.org/}, \textsuperscript{c}\url{https://www.webofknowledge.com/}}
\end{tabular}
\vspace{-0.5cm}
\end{table}

\vspace{-1.5em}
\subsubsection{Expert Interviews}
\label{sec:method-interviews}
Following the guidelines by Runeson and Höst~\cite{runeson.host_2009_guidelines}, we conduct \acp{ei} to enrich the findings of the \ac{slr} with opinions from the industry~\cite{morse_1991_approaches} 
to reduce bias from the \ac{slr}.
The empirical method consists of two parts. First, we devise a questionnaire\footnote{Questionnaire: \url{https://anonymous.4open.science/r/trustbus2022-5C26/}} with independently reviewed questions. Second, participants are recruited by contacting individuals who have co-authored journal and conference papers on \ac{sde} (e.g., \cite{cappiello.etal_2020_data}), who have worked directly or indirectly data sovereignty projects, and who come from different institutions.
Six individuals (IT project managers, computer science and security researchers) agreed to participate 
in April and May 2021. Two interviews were conducted as synchronous, semi-structured online interviews~\cite{runeson.host_2009_guidelines}, while the remaining interviews occurred via E-Mail correspondence. Participants are informed about organizational matters (e.g., study purpose, removal of personal information, right to withdraw responses). The findings are summarized before concluding the interviews to counter selection bias and avoid misinterpretation~\cite{runeson.host_2009_guidelines}.
Similar to the data analysis in the \ac{slr}, the expert interview responses are paraphrased and analyzed using content analysis. Then, the challenges are consolidated with challenges from the \ac{slr}~(Table~\ref{tab:rq1-slr-appendix}).

\subsection{Investigator Triangulation}
\label{sec:method-rq2}

\noindent \emph{\textbf{RQ2:} \rqtwo}
We apply investigator triangulation~\cite{thurmond_2001_point} to map which \acp{pet} and \acp{tet} from the technology classification (Section~\ref{sec:background}) can mitigate the \ac{sde} challenges from RQ1. The evaluation of the individual technologies was conducted by each researcher to decrease bias in the evaluation stage and contribute to the findings' internal validity~\cite{thurmond_2001_point}. First, every researcher selects one technology from the technology classification (Section~\ref{sec:background}). Second, each researcher individually goes through the challenges from RQ1 one by one, and gauges whether the technology fully or partially addresses the challenge on a theoretical or technical basis. Third, the researchers' evaluations are consolidated with one another through rigorous discussions, and recorded.
\footnote{Assessment tables: \url{https://anonymous.4open.science/r/trustbus2022-5C26/}} The final results are summarized in a technology mapping (Table~\ref{tab:rq2-mapping}).

\vspace{-0.15cm}

\section{Technology Classification}
\label{sec:background}
\vspace{-0.15cm}


\begin{table}
    \caption{Overview of \acp{pet} and \acp{tet}}
\label{tab:background-tech-description}
\centering
\scriptsize

\begin{tabular}{p{0.4em} p{0.99\textwidth}}

\toprule
\multicolumn{2}{l}{\textbf{Technology}} \\ \midrule

\multicolumn{2}{l}{\textit{Data Communication Layer}} \\ [0.15em]
 
& \textbf{Encryption.} A core cryptographic technique to send data between entities, which only intended recipients can decrypt~\cite{diffie.hellman_1976_new}. \\   \cmidrule{2-2}

& \textbf{Anonymous Routing.} As the backbone of \ac{tor}~\cite{dingledine.etal_2004_tor}, the protocol anonymously relays messages through a distributed network while being resistant to eavesdropping and traffic analysis~\cite{syverson.etal_1997_anonymous}.\\  \midrule

\multicolumn{2}{l}{\textit{Data Storage Layer}} \\[0.15em]
 & \textbf{Decentralized Storage.} 
Distributed Hash Tables (DHTs) are at the core of decentralized storage systems and can be used to store and retrieve data distributed across the nodes of a \ac{p2p} network
~\cite{palmieri.pouwelse_2014_key}. \\   \cmidrule{2-2}

 & \textbf{\ac{se}.} 
 \ac{se} supports search functionality on the server-side without decrypting data and losing data confidentiality~\cite{song.etal_2000_practical,bosch.etal_2014_survey} through \ac{sse} or \ac{peks}. \\  \midrule

 \multicolumn{2}{l}{\textit{Data Processing Layer}} \\ [0.15em]
 & \textbf{\ac{he}.} Arithmetic operations directly on ciphertext, such that only authorized entities can decrypt and access the output of the operations~\cite{lopez.farooq_2020_multilayered,chaudhary.etal_2019_analysis}. \\   \cmidrule{2-2}
 
 & \textbf{\ac{smc}.} Data exchange participants can jointly compute functions on data without revealing their data inputs to other participants. Popular implementations are based on secret-sharing~\cite{shamir_1979_how} and garbled circuits~\cite{yao_1982_protocols}.
 \\   \cmidrule{2-2}
 
 & \textbf{\ac{fl}.} Multiple participants can train machine learning models collaboratively over remote devices~\cite{konecny.etal_2015_federated,li.etal_2020_federated}, i.e., participants keep their data localized and only share local model updates with a coordinating central server. Thus, data privacy is enhanced as data never leaves the data owner's device~\cite{papadopoulos.etal_2021_privacy,li.etal_2020_federated}. \\   \cmidrule{2-2}

 & \textbf{\ac{tee}.} Secure memory areas physically isolated from the device's operating system and applications~\cite{omtp_2009_advanced,hynes.etal_2018_demonstration}.
 Unique encryption keys are associated with hardware, making software tampering as challenging as hardware tampering~\cite{munilla-garrido.etal_2021_revealing}. \\   \cmidrule{2-2}
 
 & \textbf{\ac{dp}.} Algorithms fulfilling this privacy definition enhance privacy by adding randomized noise to an analysis, such that its results are practically identical with or without the presence of an individual data subject~\cite{dwork.etal_2006_calibrating}, providing plausible deniability. \\   \cmidrule{2-2}
 
 & \textbf{k-Anonymity.} This privacy model uses syntactic building blocks (supression and generalization) to transform a dataset such that an individual cannot be distinguished from at least $k-1$ others in the dataset~\cite{samarati.sweeney_1998_protecting,sweeney_2002_kanonymity}. \\   \cmidrule{2-2}
 
 & \textbf{Pseudonymization.} Replaces identifiers with pseudonyms via encryption, hash functions, or tokenization to decreases the linkability to individuals~\cite{niu.etal_2019_achieving,sharma.etal_2018_practical}.\\  \midrule

 \multicolumn{2}{l}{\textit{Verification Layer}} \\ [0.15em]
 & \textbf{\ac{dlt}.} Distributed and tamper-proof database, where the state is stored on multiple nodes of a cryptographically secured \ac{p2p} network~\cite{butihn.etal_2020_blockchains}. 
 The state is updated on all nodes using a consensus algorithm. \\   \cmidrule{2-2}
 
 & \textbf{\ac{vc}.} \acp{vc} are sets of verifiable claims that can prove the authenticity of attributes or identities~\cite{muhle.etal_2018_survey}. The standardized digital credentials use \acp{did} and \acp{ds} to form attestation systems. \\   \cmidrule{2-2}
 
 & \textbf{\ac{zkp}.} Cryptographic protocol allowing to authenticate knowledge without revealing the knowledge itself~\cite{goldwasser.etal_1985_knowledge,goldreich.oren_1994_definitions}. 
 \acp{zkp} and can provide data authenticity, identity authenticity, and computational integrity. \\   \midrule

  \multicolumn{2}{l}{\textit{Sovereignty Layer}} \\ [0.15em]
 & \textbf{\ac{pbd}.} The seven-step guidelines can protect privacy in systems' designs by acknowledging privacy within risk management and design processes (e.g., privacy by default settings, end-to-end security of personal data)~\cite{gurses.etal_2011_engineering,cavoukian.others_2009_privacy}.  \\   \cmidrule{2-2}
 
 & \textbf{\ac{pp}.} \acp{pp} (e.g., through smart contracts~\cite{munilla-garrido.etal_2021_revealing} or on a contractual basis~\cite{cappiello.etal_2020_data}) embody privacy requirements and guidelines of data governance models~\cite{spiekermann.novotny_2015_vision}, such that different policies can be applied to different data consumers.\\ 
 
 \bottomrule
\end{tabular}

\end{table}

\emph{\acfp{pet}} \emph{manage} data through privacy principled architectures and policies or \emph{modify} data with heuristics or mathematical privacy guarantees to protect personal or sensitive information while minimally disturbing data utility~\cite{munilla-garrido.etal_2021_revealing,hes.etal_1998_privacyenhancing,trask.etal_2020_privacy}. 
In contrast, \emph{\acfp{tet}} support and improve the assessment of authenticity and integrity of data in a digital system~\cite{munilla-garrido.etal_2021_revealing}. \acp{tet} thus facilitate the assessment of trust and confidence between parties~\cite{cofta_2007_trustenhancing,josang.pope_2005_user} and ensure accountability 
and compliance~\cite{bordel.alcarria_2021_trustenhancing}.


Inspired by~\cite{danezis.etal_2014_privacy,munilla-garrido.etal_2021_revealing}, we classify \acp{pet} and \acp{tet} into five layers. The layers correspond to data communication, storage, processing, verification, and sovereignty (Table~\ref{tab:background-tech-description}). 
The \emph{data communication layer} and \emph{storage layer} address how to transfer and store data, respectively, in a secure and trustworthy manner. 
The \emph{data processing layer} includes technologies that enhance the privacy of the data input, its computation, and output~\cite{trask.etal_2020_privacy,munilla-garrido.etal_2021_revealing}. The \emph{data verification layer} is concerned with certification of identities, properties of individuals, and attributes of datasets or resources. 
Lastly, the \emph{sovereignty layer} includes mechanisms that enforce usage control and privacy protection on a policy basis~\cite{dinev.etal_2013_information}.

\vspace{-0.5cm}
\section{Challenges \& Technology Mapping}
\label{sec:findings}

\subsection{SDE Challenges (RQ1)}
\label{sec:findings-rq1}

The concept matrix (Table~\ref{tab:rq1-slr-appendix}) lists $13$ \ac{sde} challenges identified from analyzing the final $16$ literature hits and interviewing 6 experts. 
The \ac{sde} challenges are described and grouped into organizational issues, data processing and publishing, and infrastructure issues (Table~\ref{tab:rq1-challenges}).
The \emph{organizational challenges} relate to institutions' uncertainties regarding legislation, technology standards, opportunity costs of data exchange, and by extension, \ac{sde}. The \emph{data processing and publishing challenges} are primarily concerned with protecting digital subjects' personal interests and privacy in data exchange. Lastly, \emph{Infrastructure challenges} relate to data security and privacy beyond processing and publishing data. Specifically, the challenges deal with implementation issues regarding data access and usage control, strengthening trust in the infrastructure and amongst data exchange participants, as well as enforcing accountability and auditability.

\begin{table}[tb]
    \caption{Concept Matrix of Identified SDE Challenges}
\label{tab:rq1-slr-appendix}
\centering
\scriptsize
\begin{tabular}{p{0.6em} l *{14}{c}}

\toprule
 &  & \spheadingSHORT{\textbf{Managing Jurisdictions}} & \spheadingSHORT{\textbf{Missing Standards}} & \spheadingSHORT{\textbf{Reluctance}} & \spheadingSHORT{\textbf{Ensuring Data Privacy}} & \spheadingSHORT{\textbf{Ensuring Data Quality}} & \spheadingSHORT{\textbf{Ensuring Computational Privacy}} & \spheadingSHORT{\textbf{Interoperability}} & \spheadingSHORT{\textbf{Minimizing Computational Complexity}} & \spheadingSHORT{\textbf{Inter-Organizational Trust}} & \spheadingSHORT{\textbf{Cyber Security \& Trust in Infrastructure}} & \spheadingSHORT{\textbf{Data Provenance}} & \spheadingSHORT{\textbf{Data Usage \& Access Control}} & \spheadingSHORT{\textbf{Auditability}} \\ \cmidrule{3-15}
 
\multicolumn{2}{l}{\textbf{Data Source}} & \textbf{C1} & \textbf{C2} & \textbf{C3} & \textbf{C4} & \textbf{C5} & \textbf{C6} & \textbf{C7} & \textbf{C8} & \textbf{C9} & \textbf{C10} & \textbf{C11} & \textbf{C12} & \textbf{C13} \\ \midrule
 
\multicolumn{5}{l}{\textit{Systematic Literature Review}} &  &  &  &  &  &  &  &  &  &  \\
 & Panhuis~et~al.~\cite{vanpanhuis.etal_2014_systematic} & \checkmark & \checkmark & \checkmark & \checkmark & \checkmark &  & \checkmark &  & \checkmark &  &  & \checkmark &  \\ \cline{3-15} 
 & Lablans~et~al.~\cite{lablans.etal_2015_exploiting} &  &  & \checkmark &  &  & \checkmark & \checkmark &  &  &  &  &  &  \\ \cline{3-15} 
 & Ahmadian~et~al.~\cite{ahmadian.etal_2018_extending} & \checkmark &  &  & \checkmark &  &  &  & \checkmark &  & \checkmark & \checkmark &  & \checkmark \\ \cline{3-15} 
 & Bennett~et~al.~\cite{bennett.oduro-marfo_2018_global} &  &  &  & \checkmark & \checkmark &  &  &  &  & \checkmark &  & \checkmark &  \\ \cline{3-15} 
 & Brost~et~al.~\cite{brost.etal_2018_ecosystem} &  &  &  &  &  &  &  &  & \checkmark & \checkmark &  & \checkmark &  \\ \cline{3-15} 
 & Demchenko~et~al.~\cite{demchenko.etal_2018_data} &  & \checkmark &  &  &  &  &  &  & & &  & &  \\ \cline{3-15} 
 & Celik~et~al.~\cite{celik.etal_2019_curie} & \checkmark &  & \checkmark &  &  & \checkmark &  &  &  &  &  &  &  \\ \cline{3-15} 
 & Cuno~et~al.~\cite{cuno.etal_2019_data} & \checkmark & \checkmark & \checkmark &  &  &  &  &  &  &  &  &  &  \\ \cline{3-15} 
 & Otto~et~al.~\cite{otto.jarke_2019_designing} &  &  &  & \checkmark & \checkmark &  & \checkmark & \checkmark & \checkmark & \checkmark & \checkmark & \checkmark &  \\ \cline{3-15} 
 & Sarabia-Jacome~et~al.~\cite{sarabia-jacome.etal_2019_enabling} &  &  &  &  &  &  & \checkmark &  &  &  &  &  &  \\ \cline{3-15} 
 & Zrenner~et~al.~\cite{zrenner.etal_2019_usage} &  &  &  & \checkmark &  &  & \checkmark & \checkmark &  &  &  & \checkmark &  \\ \cline{3-15} 
 & Gil~et~al.~\cite{gil.etal_2020_evaluation} &  &  &  &  &  &  &  &  & \checkmark &  & \checkmark & \checkmark &  \\ \cline{3-15} 
 & Lee~et~al.~\cite{lee.etal_2020_coded} &  &  &  & \checkmark &  &  &  &  &  &  &  &  &  \\ \cline{3-15} 
 & Nast~et~al.~\cite{nast.etal_2020_workinprogress} &  &  &  &  &  & \checkmark &  &  &  &  &  & \checkmark &  \\ \cline{3-15} 
 & Andreas~et~al.~\cite{andreas.etal_2021_optimized} & \checkmark &  &  & \checkmark &  &  & \checkmark & \checkmark &  &  &  &  &  \\ \cline{3-15} 
 & Grünewald~et~al.~\cite{grunewald.pallas_2021_tilt} &  & \checkmark &  & \checkmark &  &  &  &  &  & \checkmark & \checkmark &  &  \\ \cline{3-15} 
\multicolumn{2}{l}{Count from \ac{slr}} & 5 & 4 & 4 & 8 & 3 & 3 & 6 & 4 & 4 & 5 & 4 & 7 & 1 \\ \midrule

\multicolumn{2}{l}{\textit{Expert Interviews}} &  &  &  &  &  &  &  &  &  &  &  &  &  \\
 & \interviewee{1}{int:A} [\ref{int:A}] &  &  &  &  & \checkmark &  & \checkmark &  &  &  & \checkmark & \checkmark &  \\ \cline{3-15} 
 & \interviewee{2}{int:B} [\ref{int:B}] &  &  &  &  &  &  &  &  & \checkmark &  &  & \checkmark &  \\ \cline{3-15} 
 & \interviewee{3}{int:C} [\ref{int:C}]&  &  &  &  &  & \checkmark &  &  &  &  &  & \checkmark &  \\ \cline{3-15} 
 & \interviewee{4}{int:D} [\ref{int:D}]&  & \checkmark & \checkmark & \checkmark &  & \checkmark & \checkmark &  &  &  &  & \checkmark & \checkmark \\ \cline{3-15} 
 & \interviewee{5}{int:E} [\ref{int:E}]&  &  &  & \checkmark &  & \checkmark &  &  & \checkmark &  & \checkmark & \checkmark & \checkmark \\ \cline{3-15} 
 & \interviewee{6}{int:F} [\ref{int:F}]&  & \checkmark &  &  &  &  &  &  &  &  & \checkmark & \checkmark &  \\ \cline{3-15} 
\multicolumn{2}{l}{Count from Interviews} & - & 2 & 1 & 2 & 1 & 3 & 2 & - & 2 & - & 3 & 6 & 2 \\ \midrule 

\multicolumn{2}{l}{\textbf{Total No. References}} & \textbf{5} & \textbf{6} & \textbf{5} & \textbf{10} & \textbf{4} & \textbf{6} & \textbf{8} & \textbf{4} & \textbf{6} & \textbf{5} & \textbf{7} & \textbf{13} & \textbf{3}
\\ 
\bottomrule
\end{tabular}
\vspace{-0.5cm}
\end{table}

\begin{table}
    \caption{Identified SDE Challenges (RQ1)}
\label{tab:rq1-challenges}
\centering
\scriptsize

\begin{tabular}{p{0.4em} c p{0.942\textwidth}}

    \toprule
    \multicolumn{2}{c}{\textbf{No.}} & \textbf{\ac{sde} Challenge Description}  \\\midrule
    
    \multicolumn{3}{l}{\textit{Organizational Challenges}} \\[0.15em]
    & \textbf{C1} & \textbf{Managing Jurisdictions.} Uncertainties about interpreting and implementing legal guidelines in technological infrastructures~\cite{ahmadian.etal_2018_extending} (e.g., data sharing, copyright, data ownership, or personal data) in different jurisdictions~\cite{andreas.etal_2021_optimized,celik.etal_2019_curie,vanpanhuis.etal_2014_systematic}. Varying regulations for (at times hard to distinguish) personal and anonymized datasets cause further complications 
    ~\cite{cuno.etal_2019_data,vanpanhuis.etal_2014_systematic}. \\\cmidrule{3-3} 
    
    & \textbf{C2} & \textbf{Missing Standards.} The definition and implementation of suitable technological solutions for \ac{sde} use cases lack standards~[\ref{int:B}], e.g., lack of end-to-end standards~[\ref{int:E}] for data exchange, use, and replication~\cite{cuno.etal_2019_data}, technologically translating data protection regulations~\cite{grunewald.pallas_2021_tilt,demchenko.etal_2018_data}, data formats~\cite{vanpanhuis.etal_2014_systematic}, and proving data integrity and authenticity~[\ref{int:F}]. \\\cmidrule{3-3}
    
    & \textbf{C3} & \textbf{Reluctance.} Despite trustworthy technical guarantees, organizations hesitate to engage in data exchange because the terminology \say{sharing/exchange} suggests that raw data is transferred to third parties~[\ref{int:D}]. Organizations lack incentives~\cite{vanpanhuis.etal_2014_systematic}, as the benefits of data exchange rarely outweigh high opportunity costs (e.g., privacy and (sensitive) data breaches)~\cite{lablans.etal_2015_exploiting}. \\\midrule
    
    \multicolumn{3}{l}{\textit{Data Processing and Publishing Challenges}} \\[0.15em]
    
    & \textbf{C4} & \textbf{Ensuring Data Privacy.} 
    Data privacy requires data to be manipulated and protected (e.g., minimize data disclosure, secure data storage)~\cite{bennett.oduro-marfo_2018_global,otto.jarke_2019_designing} to prevent unauthorized third party to draw conclusions about individual entities~[\ref{int:D}] and following privacy regulations (e.g., \ac{gdpr})~\cite{lee.etal_2020_coded,grunewald.pallas_2021_tilt}. 
    \\\cmidrule{3-3}
    
    & \textbf{C5} & \textbf{Ensuring Data Quality.} Privacy-sensitive data should be usable after anonymization. Thus, the challenge is to enable C4 while preserving data usability~\cite{otto.jarke_2019_designing}, i.e., to meet data quality requirements of data consumers~[\ref{int:A}], to incentivize data exchange. \\\cmidrule{3-3}

    & \textbf{C6} & \textbf{Ensuring Computational Privacy.} Computational privacy refers to keeping the metadata, and semantic views of a data transaction secret~\cite{lablans.etal_2015_exploiting}: who transferred data to whom, what algorithm was applied, and which dataset was accessed~[\ref{int:D}]. Thus, ensuring computational privacy is a challenge that extends C4 beyond the dataset's content. \\\cmidrule{3-3}
    
    & \textbf{C7} & \textbf{Interoperability.} Technological interoperability standardizes interactions between parties, e.g., authentication, authorization, or data exchange agreement protocols~\cite{otto.jarke_2019_designing}. Semantic interoperability 
    describes datasets through standardized metadata schemes stored in central metadata repositories~\cite{vanpanhuis.etal_2014_systematic} to facilitate the search of heterogeneous or non-standardized datasets or handle datasets that have been normalized differently~[\ref{int:A}]. \\\cmidrule{3-3}
    
    & \textbf{C8} & \textbf{Minimizing Computational Complexity.} Data processing and publishing needs scalable and affordable \ac{sde} implementations~\cite{zrenner.etal_2019_usage}. Data exchange requires low latency, computational complexity, and parallel processing to limit calculation time and memory usage~\cite{andreas.etal_2021_optimized} to make privacy-preserving alternatives affordable and usable~\cite{otto.jarke_2019_designing}. \\\midrule

    \multicolumn{3}{l}{\textit{Infrastructure Challenges}} \\[0.15em]
    
    & \textbf{C9} & \textbf{Inter-Organizational Trust.} Parties do not share data unless they trust the other parties~\cite{gil.etal_2020_evaluation} (e.g., platform operators). Companies are more likely to trust others if they have had contact with one another before~[\ref{int:B}], apply unenforceable, and untrackable soft agreements~[\ref{int:D}], or operate in a data ecosystem with a trusted root of trust~[\ref{int:A}]. \\\cmidrule{3-3}

    & \textbf{C10} & \textbf{Cyber Security \& Trust in Infrastructure.} \ac{sde} must ensure \textit{confidentiality}, \textit{integrity}, \textit{availability}, and \textit{resilience} of data and the infrastructure (soft- and hardware). Specifically, the infrastructure must be transparent~\cite{grunewald.pallas_2021_tilt,ahmadian.etal_2018_extending}, data communication and storage secured, and unauthorized access prohibited~\cite{brost.etal_2018_ecosystem}. \\\cmidrule{3-3}
    
    & \textbf{C11} & \textbf{Data Provenance.} For data exchange, data provenance is often enabled through blockchain-based technologies, logging, and monitoring transactions~\cite{otto.jarke_2019_designing} to establish \textit{accountability}~\cite{ahmadian.etal_2018_extending}, integrity, non-manipulation of data~[\ref{int:A}], and authentication processes~\cite{grunewald.pallas_2021_tilt,gil.etal_2020_evaluation}. Decisions include how long data is stored, what data is stored, and how privacy protection is handled~\cite{grunewald.pallas_2021_tilt}. \\\cmidrule{3-3}
    
    & \textbf{C12} & \textbf{Data Usage \& Access Control.} Legal and technological data control is lost after data is exchanged~\cite{zrenner.etal_2019_usage,cuno.etal_2019_data}, i.e., there is a high risk of unauthorized copying, redistribution, or reusing data for unintended purposes~[\ref{int:E}] by unauthorized parties~[\ref{int:F}]. Liabilities and accountability must be transferred with data to revoke access rights, specify a data storage location, intervene manually, or specify purpose- and time-based access rights to data
    ~\cite{brost.etal_2018_ecosystem}. \\\cmidrule{3-3}
    
    & \textbf{C13} & \textbf{Auditability.} Proving the legitimacy of claims about complying with data-related guidelines~\cite{ahmadian.etal_2018_extending} (e.g., proving that datasets have been anonymized by applying \acp{pet}, proving that consent and policy agreements are followed) is required for accountability. \\
    
    \bottomrule
    
\end{tabular}

\end{table}

As described in the methodology (Section~\ref{sec:method-rq1}), the individual challenges are derived through content analysis. For example, \emph{missing standards} (C2) is mentioned by three literature sources and two interviewees. 
Researchers 
highlighted the challenge of missing standardizing guidelines on data sharing~\cite{vanpanhuis.etal_2014_systematic} or faced missing standards regarding the publishing of smart city data~\cite{cuno.etal_2019_data}. The missing end-to-end standards are also mentioned~[\ref{int:D}]. More generally, Grünewald and Pallas~\cite{grunewald.pallas_2021_tilt} and Demchenko~et~al.~\cite{demchenko.etal_2018_data} 
encounter challenges of missing standardization in inter-system communication, standardization of transparency information items, and reusable open-source solutions to technically ensure that \ac{gdpr} rights are met. The technical limitations are also picked up by Interviewee 6~[\ref{int:F}]. The limitations caused by missing data format and architecture challenges were summarized into one challenge. As standardization processes typically require collaborative working groups or experts (e.g., ISO), C2 was considered an organizational challenge. The remaining challenges were derived in a similar manner.

\subsection{Technology Mapping (RQ2)}
\label{sec:findings-rq2}

As described in Section~\ref{sec:method-rq2}, \acp{pet} and \acp{tet} (Section~\ref{sec:background}) are evaluated based on their potential to mitigate the \ac{sde} challenges identified in RQ1. 
We summarize the evaluation in a \emph{technology mapping}~(Table~\ref{tab:rq2-mapping}). 
For example, \ac{dp} achieves anonymization in a dataset by adding randomized noise to the data. We find that \ac{dp} does not support any organizational challenge as it only addresses datasets' contents. Thus, challenges C1, C2, and C3 are not addressed, and the cells remain blank. Contrarily, \ac{dp} addresses data processing and publishing challenges. Specifically, the technique ensures data privacy (C4) while maintaining data quality (C5) and providing parameter that helps balancing privacy and accuracy. However, data quality is reduced as data is perturbed, meaning that C5 is only partially maintained. Furthermore, \ac{dp} does not have high computational complexity compared to other processing layer technologies and thus mitigates C8. 
\ac{dp} can affect inter-organizational trust (C9) in the peripheral and provide the parameters used for \ac{dp} to partially fulfill compliance verification (C13).

\begin{table}[tb]
    \caption{Mapping PETs and TETs to SDE Challenges (RQ2)}
\label{tab:rq2-mapping}

\centering
\scriptsize
\begin{tabular}{p{0.4em} >{\raggedright}p{3.3cm}cccccccccccccc}
\toprule

&  & \spheadingSHORT{\textbf{Managing Jurisdictions}} & \spheadingSHORT{\textbf{Missing Standards}} & \spheadingSHORT{\textbf{Reluctance}} & \spheadingSHORT{\textbf{Ensuring Data Privacy}} & \spheadingSHORT{\textbf{Ensuring Data Quality}} & \spheadingSHORT{\textbf{Ensuring Computational Privacy}} & \spheadingSHORT{\textbf{Interoperability}} & \spheadingSHORT{\textbf{Minimizing Computational Complexity}} & \spheadingSHORT{\textbf{Inter-Organizational Trust}} & \spheadingSHORT{\textbf{Cyber Security \& Trust in Infrastructure}} & \spheadingSHORT{\textbf{Data Provenance}} & \spheadingSHORT{\textbf{Data Usage \& Access Control}} & \spheadingSHORT{\textbf{Auditability}} \\ \cmidrule{3-15}
 
 \multicolumn{2}{l}{\textbf{Technology}} & \textbf{C1} & \textbf{C2} & \textbf{C3} & \textbf{C4} & \textbf{C5} & \textbf{C6} & \textbf{C7} & \textbf{C8} & \textbf{C9} & \textbf{C10} & \textbf{C11} & \textbf{C12} & \textbf{C13} \\ \midrule

 \multicolumn{15}{l}{\textit{Data Communication Layer}} \\ 
 
& Encryption & & & & & & & & & & \checkmark & $\sbullet$ & $\sbullet$ & \\   \cline{3-15} 
& Anonymous Routing & & & & & & $\sbullet$ & & & & $\sbullet$ & & & \\  \midrule

\multicolumn{15}{l}{\textit{Data Storage Layer}} \\ 
 & Decentralized Storage & & & & & \checkmark & & & & $\sbullet$ & $\sbullet$ & \checkmark & & \\   \cline{3-15} 
 & Searchable Encryption & & & & $\sbullet$ & \checkmark & \checkmark & $\sbullet$ & & $\sbullet$ & \checkmark & & \checkmark & \\  \midrule

 \multicolumn{15}{l}{\textit{Data Processing Layer}} \\ 
 & Homomorphic Encryption & & & & \checkmark & \checkmark & $\sbullet$ & & & $\sbullet$ & \checkmark & & $\sbullet$ & \\   \cline{3-15} 
 & Secure Multiparty Computation & & & & \checkmark & \checkmark & $\sbullet$ & & & $\sbullet$ & $\sbullet$ & & \checkmark & $\sbullet$ \\   \cline{3-15} 
 & Federated Learning & & & & \checkmark & \checkmark & $\sbullet$ & & $\sbullet$ & $\sbullet$ & $\sbullet$ & & \checkmark & \\   \cline{3-15} 
 & Trusted Execution Environment & & & & \checkmark & \checkmark & \checkmark & & $\sbullet$ & & \checkmark & $\sbullet$ & & \\   \cline{3-15} 
 & Differential Privacy & & & & \checkmark & $\sbullet$ & & & \checkmark & $\sbullet$ & & & & $\sbullet$ \\   \cline{3-15} 
 & k-Anonymity & & & & \checkmark & $\sbullet$ & & & $\sbullet$ & $\sbullet$ & & & & $\sbullet$ \\   \cline{3-15} 
 & Pseudonymization & & & & $\sbullet$ & \checkmark & & & \checkmark & & & & & \\  \midrule

 \multicolumn{15}{l}{\textit{Verification Layer}} \\ 
 & Distributed Ledger &  &  &  &  & $\sbullet$ &  &  &$\sbullet$ &\checkmark &$\sbullet$ &\checkmark &  & $\sbullet$ \\   \cline{3-15} 
 & Verifiable Credential & & \checkmark & & $\sbullet$ & & $\sbullet$ & \checkmark & \checkmark & \checkmark & $\sbullet$ & \checkmark & $\sbullet$ & \\   \cline{3-15} 
 & Zero-Knowledge Proof & & & & \checkmark & \checkmark & $\sbullet$ & & $\sbullet$ & \checkmark & $\sbullet$ & \checkmark & & \checkmark \\   \midrule

  \multicolumn{15}{l}{\textit{Sovereignty Layer}} \\ 
 & Privacy-by-Design & & $\sbullet$ & $\sbullet$ & \textasteriskcentered & \textasteriskcentered & \textasteriskcentered & & \textasteriskcentered & \textasteriskcentered & \textasteriskcentered & \textasteriskcentered & \textasteriskcentered & \textasteriskcentered \\   \cline{3-15} 
 & Privacy Policies & $\sbullet$ & & $\sbullet$ & \textasteriskcentered & \textasteriskcentered & \textasteriskcentered & \textasteriskcentered & \textasteriskcentered & \textasteriskcentered & \textasteriskcentered & \textasteriskcentered & \checkmark & \textasteriskcentered \\ \midrule

 \multicolumn{2}{l}{\textbf{Total Count (excl. \textasteriskcentered)}} & \textbf{1} & \textbf{2} & \textbf{2} & \textbf{10} & \textbf{11} & \textbf{8} & \textbf{2} & \textbf{8} & \textbf{10} & \textbf{11} & \textbf{5} & \textbf{7} & \textbf{5} \\
 \bottomrule
 
 \multicolumn{15}{l}{\checkmark~: addressed technically and theoretically} \\
 \multicolumn{15}{l}{$\sbullet$~~~: addressed technically or theoretically}\\
  \multicolumn{15}{l}{\textasteriskcentered~~: characteristics depend on the selected technologies to fulfill the requirements}
\end{tabular}

\vspace{-0.6cm}

\end{table}

\vspace{-0.2cm}

\section{Discussion}
\label{sec:discussion}
\vspace{-0.2cm}

The following section draws key findings (KF) and implications from the \ac{sde} challenges (RQ1) and the technology mapping (RQ2), outlines the study's limitations, and proposes research questions for future work.

\vspace{-1em}
\paragraph{KF1:}\emph{The final literature hits can be categorized into three thematic research streams: security and challenges of data exchange, \acp{ids}, and design of sovereignty layer technologies. The most recent research stream focuses on sovereignty layer technologies, suggesting an upward trend in the research area of \ac{sde}.}
The first research stream more generally focuses on the security and/or challenges of data exchange ($6$ studies, published 2014--2021) and the \ac{ids} stream describes implementations and use cases for the  \ac{idsa}'s data governance framework ($5$ studies, published 2018--2020), e.g., urban~\cite{cuno.etal_2019_data} or seaport data spaces~\cite{sarabia-jacome.etal_2019_enabling}. 
The third research stream focuses on technologies of the sovereignty layer (e.g., privacy languages, privacy policies, consent frameworks) and contains $5$ publications since 2019. 

\vspace{-1em}
\paragraph{KF2:}\emph{No single \ac{pet} or \ac{tet} addresses all \ac{sde} challenges, suggesting that \acp{pet} or \acp{tet} must be carefully chosen, combined, further developed, or complemented with new techniques.} 
Choosing suitable technologies for \ac{sde} use cases is essential. For example, data provenance (C11) requires logging and surveillance to establish accountability. While trackability and traceability (e.g., logging) can support C11, auditability (C13), and inter-organizational trust (C9)~\cite{otto.jarke_2019_designing}, it impedes ensuring data privacy (C4) and ensuring computational privacy (C6) and could increase the reluctance (C3) of data exchange participants. Practitioners must therefore carefully select their primary challenge.

\vspace{-1em}
\paragraph{KF3:}\emph{A large portion of \acp{pet} and \acp{tet} address data processing and publishing challenges (C4--C8), and barely address organizational challenges (C1--C3). This suggests that the research field of \ac{sde} is still maturing and has not yet established best practices.}
While data processing and publishing challenges are the core technical concepts of \ac{sde}, organizational challenges depict issues that arise in productive \ac{sde} systems. However, as best practices for data processing and publishing challenges are still in development~[\ref{int:B},\ref{int:E}], there are no productive \ac{sde} systems, rendering the organizational challenges redundant and peripheral. We deduce that research on organizational challenges will likely gain importance once data processing and publishing challenges have been thoroughly explored and state-of-the-art solutions are presented. The first indications are the growing interest in policy and enforcement research~\cite{cuno.etal_2019_data} (e.g., Gaia-X~\cite{bmwi_2020_project}, \ac{idsa}~\cite{otto.etal_2017_reference}) to leverage data sharing while protecting sensitive data~[\ref{int:D}].

\vspace{-1em}
\paragraph{KF4:}\emph{Data usage and access control (C12) is the most critical challenge for researchers and experts.}
Mechanisms must inhibit the unauthorized redistribution of data, revoke access rights, and transfer liabilities with data~[\ref{int:E}]. The fundamental technical problem is to introduce data sharing without allowing third parties to use privacy-sensitive or confidential data for non-designated purposes~[\ref{int:D}]. Access rights, remain a major technological challenge given the fluid nature of data~\cite{cuno.etal_2019_data,zrenner.etal_2019_usage} and should be addressed by researchers.
Thus, more attention should be given to implement C12 (e.g., \ac{fl} can enable data usage and access control because data is never shared with a third party, \ac{smc} can control data usage and access control by exchanging encryption keys regularly).

\vspace{-1em}
\paragraph{KF5:}\emph{Given that the research field of \ac{sde} is still maturing (KF3), challenges that are currently rarely mentioned by researchers and practitioners, i.e., auditability (C13), managing jurisdictions (C1), and minimizing computational complexity  (C8), are likely to gain importance as \acp{pet} and \acp{tet} face real world barriers.}
Although the relevance of C1, C8, and C13 is recognized~\cite{ahmadian.etal_2018_extending,vanpanhuis.etal_2014_systematic}, the challenges remain on the peripheral of \acp{pet} and \acp{tet} research. Languages to interpret jurisdictions must be refined~\cite{gerl.meier_2019_privacy,grunewald.pallas_2021_tilt} and the computational complexity of technologies must be managed to become established technologies~\cite{andreas.etal_2021_optimized,munilla-garrido.etal_2021_revealing}. We thus anticipate an increased need for research on C1, C8, and C13.

\vspace{-1.5em}
\subsubsection{Limitations} 

Even though a rigorous research design and process was adopted, there are several limitations to \acp{slr} and interviews that may undermine the effectiveness of the conceptual framework. 
Even though we followed guidelines for the respective methodologies~\cite{runeson.host_2009_guidelines,vombrocke.etal_2009_reconstructing,webster.watson_2002_analyzing}, the findings risked being incomplete, biased, or inaccurate. 
To minimize bias and strengthen the findings, we defined inclusion and exclusion criteria ex-ante and applied triangulation. 
Additionally, the sample of interviewees was low and with a varying levels of expertise. However, we employed the \acp{ei} as supporting evidence for the \ac{slr} and not as standalone findings.
Furthermore, the latter limitation was addressed by sharing the questionnaire with interviewees prior to the interview for preparation. 
Lastly, we note that, while we applied investigator triangulation to map technologies with \ac{sde} challenges, we have not tested the usability of the mapping.

\vspace{-1.5em}
\subsubsection{Future Work}
\label{sec:future-work}

We encourage researchers and practitioners to use the concept matrix (Table~\ref{tab:rq1-slr-appendix}) and technology mapping (Table~\ref{tab:rq2-mapping}) as a starting point to locate \ac{sde} challenges, identify research gaps, or choose suitable \acp{pet} or \acp{tet} for \ac{sde} use cases.
For example, interoperability (C7) is only addressed by \ac{se}, \ac{vc}, and \ac{pp}. \emph{What solution propositions exist to mitigate interoperability challenges of data and system architectures on a large scale?} Similarly, \emph{how can the auditability challenge be mitigated without compromising data privacy and computational privacy goals?}
Additionally, researchers can refine the \ac{sde} challenges and the technology mapping to refine the body of knowledge. 
Potential research questions could read: \emph{Which \acp{pet} and \acp{tet} have the potential to mitigate \emph{or worsen} \ac{sde} challenges?} 
\emph{What are best practices to implement \acp{pet} and \acp{tet} in \ac{sde} use cases?} \emph{Which \acp{pet} and/or \acp{tet} act as complements or supplements in the context of \ac{sde}?} Furthermore, \ac{pet} or \ac{tet} experts can revise and refine the technology mapping.
Finally, practitioners can use the existing technology mapping as a starting point to implement \ac{sde} use cases, and thereby help evaluate the usability of the technology mapping in a real-world context.

\vspace{-0.2cm}

\section{Related Work}
\label{sec:related-work}
\vspace{-0.2cm}

Several surveys presented an overview of \acp{pet} and \acp{tet}~\cite{adams_2021_introduction,deswarte.melchor_2006_current,senicar.etal_2003_privacyenhancing} or taxonomies with the technologies' qualities (e.g., cryptographic foundation, data handling requirements)~\cite{heurix.etal_2015_taxonomy,goldberg.etal_1997_privacyenhancing}, adoption challenges (e.g., legal, social, technical, and economic)~\cite{kaaniche.etal_2020_privacy,zoll.etal_2021_privacysensitive}, or \acp{pet} in different contexts (e.g., blockchain, personalization)~\cite{javed.etal_2021_petchain,parra-arnau.etal_2015_privacyenhancing}.
However, while these surveys presented extensive research to understand and contextualize \acp{pet}, they did not focus on data exchange. 
Only few studies investigated \acp{pet} and \acp{tet} in the context of data exchange~\cite{munilla-garrido.etal_2021_revealing} and data flow~\cite{pennekamp.etal_2019_dataflow}.
Although the studies presented useful tables mapping \acp{pet} and \acp{tet} against characteristics (e.g., authenticity, confidentiality), the privacy-oriented mappings did not focus on supporting the application of \acp{pet} and \acp{tet}.
Similarly, a data sovereignty challenge-oriented mapping~\cite{lauf.etal_2022_linking} only superficially presented solution approaches instead of concrete \acp{pet} and \acp{tet}.

Others focused on the application of \acp{pet} and \acp{tet}. There are handbooks describing how to design privacy-preserving software agents~\cite{borking.raab_2001_laws}, legally implement privacy requirements~\cite{danezis.etal_2014_privacy}, or how to adopt \acp{pet} (without including \acp{tet}) using a question-based flowchart~\cite{cdei_2021_privacy}.
Similarly, there are overviews of business use cases for which \acp{pet} can be applied~\cite{clarke_2008_business,jaatun.etal_2012_privacy,fischer-hbner.berthold_2017_chapter}. In contrast, other researchers~\cite{munilla-garrido.etal_2021_exploring} outlined which \acp{pet} and \acp{tet} can address challenges of value chain use cases. However, the use case-oriented mappings did not supports the implementation of \acp{pet} and \acp{tet}. 
More concretely, Papadopoulos~et~al.~\cite{papadopoulos.etal_2021_privacy} presented a use case that implements \ac{fl} to meet the privacy and trust requirements of the involved participants. They demonstrated that the challenges of \ac{sde} are addressable, but did not provide a framework to support the implementation of similar endeavors.
Furthermore, there exist data ecosystem reference architectures (e.g., Gaia-X~\cite{bmwi_2020_project} or \ac{idsa}~\cite{otto.etal_2017_reference}) and policy frameworks~\cite{zrenner.etal_2019_usage}. Although the reference architectures have presented holistic solution propositions to \ac{sde} challenges, the data ecosystems are not yet practicable. 

Overall, studies either (i) investigated the potentials of \acp{pet} and \acp{tet} without a specific focus on \ac{sde}, or (ii) addressed \ac{sde} challenges with an individual \ac{pet} or \ac{tet}. No study proposes a technology mapping to help practitioners and researchers identify suitable \acp{pet} and \acp{tet} when implement \ac{sde} use cases. However, this work is necessary to support researchers and practitioners in understanding and integrating the technologies in practice.
\section{Conclusion}
\label{sec:conclusion}

With this study, we structure the landscape of \ac{sde} challenges and identify suitable mitigating technologies, thereby guiding the implementation of \ac{sde} use cases and informing researchers of potential future research areas.
A two-pronged approach was pursued. First, we identified $13$ \ac{sde} challenges through a \ac{slr} and \acp{ei} (Table~\ref{tab:rq1-challenges}). Second, we proposed \acp{pet} and \acp{tet} to mitigate the identified \ac{sde} challenges using investigator triangulation, summarized in a technology mapping (Table~\ref{tab:rq2-mapping}). 
The technology mapping synthesizes previous research in an actionable way for practitioners and researchers by presenting a challenge-oriented approach that supports the identification of appropriate \acp{pet} and \acp{tet} to tackle \ac{sde} challenges -- regardless of the use case. No single technology mitigates all \ac{sde} challenges, indicating that \acp{pet} and \acp{tet} can be combined, further investigated by researchers, or complemented with new solutions. In particular, we suggest focusing on access control and jointly facilitating auditability and data privacy.


%

\bibliographystyle{splncs04}
\bibliography{References}

\begin{thebibliography}{10}
\providecommand{\url}[1]{\texttt{#1}}
\providecommand{\urlprefix}{URL }
\providecommand{\doi}[1]{https://doi.org/#1}

\bibitem{adams_2021_introduction}
Adams, C.: Introduction to Privacy Enhancing Technologies: {A}
  Classification-Based Approach to Understanding {PETs}. Springer, Cham,
  Switzerland (2021)

\bibitem{ahmadian.etal_2018_extending}
Ahmadian, A.S., Jürjens, J., Strüber, D.: Extending model-based privacy
  analysis for the industrial data space by exploiting privacy level
  agreements. In: Proceedings of the 33rd annual {ACM} symposium on applied
  computing. pp. 1142--1149 (2018)

\bibitem{andreas.etal_2021_optimized}
Andreas, A., {et al.}: Towards an optimized security approach to {IoT} devices
  with confidential healthcare data exchange. Multimedia Tools and Applications
   \textbf{80}(20),  31435--31449 (2021). \doi{10.1007/s11042-021-10827-x}

\bibitem{bastiaansen.etal_2020_usercentric}
Bastiaansen, H.J., Kollenstart, M., Dalmolen, S., van Engers, T.M.:
  User-centric network-model for data control with interoperable legal data
  sharing artefacts: {Improved} data sovereignty, trust and security for
  enhanced adoption in interorganizational and supply chain in applications.
  In: 24th Pacific Asia Conference on Information Systems. pp. 1--14. AIS,
  Dubai, UAE (2020)

\bibitem{bennett.oduro-marfo_2018_global}
Bennett, C., Oduro-Marfo, S.: {GLOBAL} privacy protection: {Adequate} laws,
  accountable organizations and/or data localization? In: 2018 ACM
  International Joint Conference on Pervasive and Ubiquitous Computing. pp.
  880--890 (2018)

\bibitem{bmwi_2020_project}
{BMWi}: Project {GAIA}-{X}: {A} federated data infrastructure as the cradle of
  a vibrant european ecosystem. Tech. rep., Federal Ministry for Economic
  Affairs and Energy (BMWi), Berlin, Germany (2020)

\bibitem{bordel.alcarria_2021_trustenhancing}
Bordel, B., Alcarria, R.: Trust-enhancing technologies: {Blockchain}
  mathematics in the context of {Industry} 4.0. In: Advances in mathematics for
  industry 4.0, pp. 1--22. Academic Press, Amsterdam, Netherlands (2021)

\bibitem{borking.raab_2001_laws}
Borking, J.J., Raab, C.D.: Laws, {PETs} and other technologies for privacy
  protection. Journal of Information, Law and Technology  \textbf{1},  1--14
  (2001)

\bibitem{brost.etal_2018_ecosystem}
Brost, G., Huber, M., Weiß, M., Protsenko, M., Schütte, J., Wessel, S.: An
  ecosystem and {IoT} device architecture for building trust in the industrial
  data space. In: Proceedings of the 4th {ACM} workshop on cyber-physical
  system security. pp. 39--50. ACM, Incheon, Republic of Korea (2018)

\bibitem{butihn.etal_2020_blockchains}
Butijn, B.J., Tamburri, D.A., Heuvel, W.J.v.d.: Blockchains: A systematic
  multivocal literature review. ACM Computing Surveys  \textbf{53}(3),  1--37
  (2020)

\bibitem{bosch.etal_2014_survey}
Bösch, C., Hartel, P., Jonker, W., Peter, A.: A survey of provably secure
  searchable encryption. ACM Computer Surveys  \textbf{47}(2),  1--51 (2014)

\bibitem{cappiello.etal_2020_data}
Cappiello, C., Gal, A., Jarke, M., Rehof, J.: Data ecosystems: {Sovereign} data
  exchange among organizations. Dagstuhl Reports  \textbf{9}(9),  66--134
  (2020)

\bibitem{cavoukian.others_2009_privacy}
Cavoukian, A.: Privacy by design: {The} 7 foundational principles. Tech. rep.,
  Information and privacy commissioner of Ontario, Canada (2009)

\bibitem{celik.etal_2019_curie}
Celik, Z.B., Acar, A., Aksu, H., Sheatsley, R., McDaniel, P., Uluagac, A.S.:
  Curie: {Policy}-based secure data exchange. In: Proceedings of the ninth
  {ACM} conference on data and application security and privacy. pp. 121--132.
  ACM (2019)

\bibitem{cdei_2021_privacy}
{Centre for Data Ethics and Innovation (CDEI)}: Privacy enhancing technologies
  adoption guide (2021), \url{https://cdeiuk.github.io/pets-adoption-guide/}

\bibitem{chaudhary.etal_2019_analysis}
Chaudhary, P., Gupta, R., Singh, A., Majumder, P.: {Analysis and Comparison of
  Various Fully Homomorphic Encryption Techniques}. In: 2019 International
  Conference on Computing, Power and Communication Technologies. pp. 58--62
  (2019)

\bibitem{clarke_2008_business}
Clarke, R.: Business cases for privacy-enhancing technologies. In: Computer
  security, privacy and politics. IRM Press, New York, USA (2008)

\bibitem{cofta_2007_trustenhancing}
Cofta, P.: Trust-enhancing technologies. In: Trust, Complexity and Control, pp.
  187--205. John Wiley and Sons, Ltd, West Sussex, England, UK (2007)

\bibitem{cuno.etal_2019_data}
Cuno, S., Bruns, L., Tcholtchev, N., Lämmel, P., Schieferdecker, I.: Data
  {Governance} and {Sovereignty} in {Urban} {Data} {Spaces} {Based} on
  {Standardized} {ICT} {Reference} {Architectures}. Data  \textbf{4}(1),  1--24
  (2019). \doi{10.3390/data4010016}

\bibitem{danezis.etal_2014_privacy}
Danezis, G., {et al.}: Privacy and {Data} {Protection} by {Design} - {From}
  {Policy} to {Engineering} (2014). \doi{10.48550/ARXIV.1501.03716}

\bibitem{demchenko.etal_2018_data}
Demchenko, Y., de~Laat, C., Los, W.: {Data as Economic Goods: Definitions,
  Properties, Challenges, Enabling Technologies for Future Data Markets}. {ITU
  Journal: ICT Discoveries}  \textbf{1}(2),  1--10 (2018).
  \doi{10.5281/zenodo.2483185}

\bibitem{deswarte.melchor_2006_current}
Deswarte, Y., Melchor, C.: {Current and Future Privacy Enhancing Technologies
  for the Internet}. Annales des Télécommunications  \textbf{61},  399--417
  (2006)

\bibitem{diffie.hellman_1976_new}
Diffie, W., Hellman, M.: New directions in cryptography. IEEE Transactions on
  Information Theory  \textbf{22}(6),  644--654 (1976)

\bibitem{dinev.etal_2013_information}
Dinev, T., Xu, H., Smith, J.H., Hart, P.: Information privacy and correlates:
  An empirical attempt to bridge and distinguish privacy-related concepts.
  European Journal of Information Systems  \textbf{22}(3),  295--316 (2013)

\bibitem{dingledine.etal_2004_tor}
Dingledine, R., Mathewson, N., Syverson, P.: Tor: The {Second-Generation} onion
  router. In: Proceedings of the 13th USENIX Security Symposium. pp. 1--17
  (2004)

\bibitem{dwork.etal_2006_calibrating}
Dwork, C., McSherry, F., Nissim, K., Smith, A.: Calibrating noise to
  sensitivity in private data analysis. In: Theory of Cryptography. pp.
  265--284. Springer (2006)

\bibitem{fischer-hbner.berthold_2017_chapter}
Fischer-Hbner, S., Berthold, S.: Privacy-enhancing technologies. In: Computer
  and Information Security Handbook, pp. 759--778. Morgan Kaufmann, 3rd edn.
  (2017)

\bibitem{gerl.meier_2019_privacy}
Gerl, A., Meier, B.: Privacy in the {Future} of {Integrated} {Health} {Care}
  {Services}-{Are} {Privacy} {Languages} the {Key}? In: 2019 {International}
  {Conference} on {Wireless} and {Mobile} {Computing}, {Networking} and
  {Communications}. pp. 312--317. IEEE (2019)

\bibitem{gil.etal_2020_evaluation}
Gil, G., Arnaiz, A., Diez, F.J., Higuero, M.V.: Evaluation {Methodology} for
  {Distributed} {Data} {Usage} {Control} {Solutions}. In: 2020 {Global}
  {Internet} of {Things} {Summit}. pp.~1--6. IEEE, Dublin, Ireland (2020)

\bibitem{goldberg.etal_1997_privacyenhancing}
Goldberg, I., Wagner, D., Brewer, E.: Privacy-enhancing technologies for the
  {Internet}. In: Proceedings {IEEE} {COMPCON} 1997. pp. 103--109. IEEE (1997)

\bibitem{goldreich.oren_1994_definitions}
Goldreich, O., Oren, Y.: Definitions and properties of zero-knowledge proof
  systems. Journal of Cryptology  \textbf{7}(1),  1--32 (1994)

\bibitem{goldwasser.etal_1985_knowledge}
Goldwasser, S., Micali, S., Rackoff, C.: The knowledge complexity of
  interactive proof-systems. In: Proceedings of the seventeenth annual {ACM}
  symposium on theory of computing. pp. 291--304. ACM, Rhode Island, USA (1985)

\bibitem{grunewald.pallas_2021_tilt}
Grünewald, E., Pallas, F.: {TILT}: {A} {GDPR}-{Aligned} transparency
  information language and toolkit for practical privacy engineering. In:
  Proceedings of the 2021 {ACM} Conference on Fairness, Accountability, and
  Transparency. pp. 636--646. ACM, Virtual Event, Canada (2021).
  \doi{10.1145/3442188.3445925}

\bibitem{gurses.etal_2011_engineering}
Gürses, S., Troncoso, C., Diaz, C.: Engineering privacy by design. In:
  Conference on computers, privacy \& data protection. pp. 1--21. CPDP
  Conferences (2011)

\bibitem{hes.etal_1998_privacyenhancing}
Hes, R., Borking, J.J. (eds.): Privacy-Enhancing Technologies: The Path to
  Anonymity. Registratiekamer, The Hague, Netherlands, revised edn. (1998)

\bibitem{heurix.etal_2015_taxonomy}
Heurix, J., Zimmermann, P., Neubauer, T., Fenz, S.: A taxonomy for privacy
  enhancing technologies. Computers \& Security  \textbf{53},  1--17 (2015)

\bibitem{hynes.etal_2018_demonstration}
Hynes, N., Dao, D., Yan, D., Cheng, R., Song, D.: A demonstration of sterling:
  {A} privacy-preserving data marketplace. Proceedings of the VLDB Endowment
  \textbf{11}(12),  2086--2089 (2018). \doi{10.14778/3229863.3236266}

\bibitem{jaatun.etal_2012_privacy}
Jaatun, M., T{\o}ndel, I.A., Bernsmed, K., Nyre, {\AA}.: Privacy enhancing
  technologies for information control. In: Privacy {Protection} {Measures} and
  {Technologies} in {Business} {Organizations}: {Aspects} and {Standards}, pp.
  1--31. IGI Global (2012)

\bibitem{javed.etal_2021_petchain}
Javed, I.T., Alharbi, F., Margaria, T., Crespi, N., Qureshi, K.N.: {PETchain}:
  {A} blockchain-based privacy enhancing technology. IEEE Access: Practical
  Innovations, Open Solutions  \textbf{9},  41129--41143 (2021)

\bibitem{josang.pope_2005_user}
J{\o}sang, A., Pope, S.: User {Centric} {Identity} {Management}. In:
  Proceedings of {AusCERT} 2005. pp. 1--13. AusCERT, Brisbane, Australia (2005)

\bibitem{kaaniche.etal_2020_privacy}
Kaaniche, N., Laurent, M., Belguith, S.: Privacy enhancing technologies for
  solving the privacy-personalization paradox: {Taxonomy} and survey. Journal
  of Network and Computer Applications  \textbf{171},  1--32 (2020)

\bibitem{konecny.etal_2015_federated}
Konečný, J., McMahan, B., Ramage, D.: Federated optimization: Distributed
  optimization beyond the datacenter (2015)

\bibitem{lablans.etal_2015_exploiting}
Lablans, M., Kadioglu, D., Muscholl, M., Ückert, F.: Exploiting distributed,
  heterogeneous and sensitive data stocks while maintaining the owner’s data
  sovereignty. Methods of information in medicine  \textbf{54}(04),  346--352
  (2015)

\bibitem{lauf.etal_2022_linking}
Lauf, F., {et al.}: Linking {Data} {Sovereignty} and {Data} {Economy}:
  {Arising} {Areas} of {Tension}. In: Wirtschaftsinformatik 2022 {Proceedings}.
  pp. 1--18. AIS (2022)

\bibitem{lee.etal_2020_coded}
Lee, A.R., Kim, M.G., Won, K.J., Kim, I.K., Lee, E.: Coded {Dynamic} {Consent}
  framework using blockchain for healthcare information exchange. In: 2020
  {IEEE} {International} {Conference} on {Bioinformatics} and {Biomedicine}.
  pp. 1047--1050 (2020)

\bibitem{li.etal_2020_federated}
Li, T., Sahu, A.K., Talwalkar, A., Smith, V.: Federated learning: Challenges,
  methods, and future directions. IEEE Signal Processing Magazine
  \textbf{37}(3),  50--60 (2020)

\bibitem{lopez.farooq_2020_multilayered}
López, D., Farooq, B.: A multi-layered blockchain framework for smart mobility
  data-markets. Transportation Research Part C: Emerging Technologies
  \textbf{111},  588--615 (2020). \doi{10.1016/j.trc.2020.01.002}

\bibitem{meuser.nagel_2009_expert}
Meuser, M., Nagel, U.: The expert interview and changes in knowledge
  production. In: Interviewing experts, pp. 17--42. Palgrave Macmillan UK,
  London (2009)

\bibitem{morse_1991_approaches}
Morse, J.M.: Approaches to {{Qualitative-Quantitative Methodological
  Triangulation}}. Nursing Research  \textbf{40}(2),  120--123 (1991)

\bibitem{munilla-garrido.etal_2021_exploring}
Munilla~Garrido, G., Schmidt, K., Harth-Kitzerow, C., Luckow, A., Matthes, F.:
  Exploring privacy-enhancing technologies in the automotive value chain. In:
  2021 {IEEE} international conference on big data. pp.~1--8. IEEE, Orlando,
  USA (2021)

\bibitem{munilla-garrido.etal_2021_revealing}
Munilla~Garrido, G., Sedlmeir, J., Uluda{\u g}, {\" O}., Alaoui, I.S., Luckow,
  A., Matthes, F.: Revealing the landscape of privacy-enhancing technologies in
  the context of data markets for the {IoT}: {A} systematic literature review
  (2021)

\bibitem{muhle.etal_2018_survey}
Mühle, A., Grüner, A., Gayvoronskaya, T., Meinel, C.: A survey on essential
  components of a self-sovereign identity. Computer Science Review
  \textbf{30},  80--86 (2018)

\bibitem{nast.etal_2020_workinprogress}
Nast, M., {et al.}: Work-in-{Progress}: {Towards} an {International} {Data}
  {Spaces} {Connector} for the {Internet} of {Things}. In: 2020 16th {IEEE}
  {International} {Conference} on {Factory} {Communication} {Systems}.
  pp.~1--4. IEEE, Porto, Portugal (2020)

\bibitem{niu.etal_2019_achieving}
Niu, C., Zheng, Z., Wu, F., Gao, X., Chen, G.: Achieving data truthfulness and
  privacy preservation in data markets. IEEE Transactions on Knowledge and Data
  Engineering  \textbf{31}(1),  105--119 (2019).
  \doi{10.1109/TKDE.2018.2822727}

\bibitem{omtp_2009_advanced}
{OMTP}: {Advanced trusted environment}. Tech. rep., OMTP Limited (2009)

\bibitem{otto.etal_2017_reference}
Otto, B., {et al.}: Reference architecture model for the industrial data space.
  Tech. rep., Fraunhofer Gesellschaft (2017)

\bibitem{otto.jarke_2019_designing}
Otto, B., Jarke, M.: Designing a multi-sided data platform: findings from the
  {International} {Data} {Spaces} case. Electronic Markets  \textbf{29}(4),
  561--580 (2019)

\bibitem{palmieri.pouwelse_2014_key}
Palmieri, P., Pouwelse, J.: Key management for onion routing in a true peer to
  peer setting. In: Advances in information and computer security. pp. 62--71
  (2014)

\bibitem{vanpanhuis.etal_2014_systematic}
van Panhuis, W.G., {et al.}: A systematic review of barriers to data sharing in
  public health. BMC Public Health  \textbf{14} (2014).
  \doi{10.1186/1471-2458-14-1144}

\bibitem{papadopoulos.etal_2021_privacy}
Papadopoulos, P., Abramson, W., Hall, A.J., Pitropakis, N., Buchanan, W.J.:
  Privacy and trust redefined in federated machine learning. Machine Learning
  and Knowledge Extraction  \textbf{3}(2),  333--356 (2021)

\bibitem{parra-arnau.etal_2015_privacyenhancing}
Parra-Arnau, J., Rebollo-Monedero, D., Forné, J.: Privacy-enhancing
  technologies and metrics in personalized information systems. In: Advanced
  research in data privacy, pp. 423--442. Springer, Cham, Switzerland (2015)

\bibitem{pennekamp.etal_2019_dataflow}
Pennekamp, J., {et al.}: Dataflow challenges in an internet of production: {A}
  security \& privacy perspective. In: Proceedings of the {ACM} {Workshop} on
  {Cyber}-{Physical} {Systems} {Security} \& {Privacy}. pp. 27--38. ACM,
  London, UK (2019)

\bibitem{runeson.host_2009_guidelines}
Runeson, P., Höst, M.: Guidelines for conducting and reporting case study
  research in software engineering. Empirical software engineering
  \textbf{14}(2),  131--164 (2009)

\bibitem{samarati.sweeney_1998_protecting}
Samarati, P., Sweeney, L.: Protecting privacy when disclosing information:
  k-anonymity and its enforcement through generalization and suppression. Tech.
  rep., Data Privacy Lab (1998)

\bibitem{sarabia-jacome.etal_2019_enabling}
Sarabia-Jacome, D., Lacalle, I., Palau, C.E., Esteve, M.: Enabling {Industrial}
  {Data} {Space} {Architecture} for {Seaport} {Scenario}. In: 2019 {IEEE} 5th
  {World} {Forum} on {Internet} of {Things}. pp. 101--106. IEEE, Limerick,
  Ireland (2019)

\bibitem{schmidt.etal_2021_exploitative}
Schmidt, K., Ullrich, A., Eigelshoven, F.: From exploitative structures towards
  data subject-inclusive personal data markets – a systematic literature
  review. In: Proceedings of the 29th european conference on information
  systems (2021)

\bibitem{senicar.etal_2003_privacyenhancing}
Seničar, V., Jerman-Blažič, B., Klobučar, T.: Privacy-enhancing
  {Technologies}—approaches and development. Computer Standards \& Interfaces
   \textbf{25}(2),  147--158 (2003). \doi{10.1016/S0920-5489(03)00003-5}

\bibitem{shamir_1979_how}
Shamir, A.: How to share a secret. Communications of the ACM  \textbf{22}(11),
  612--613 (1979). \doi{10.1145/359168.359176}

\bibitem{sharma.etal_2018_practical}
Sharma, S., Chen, K., Sheth, A.: Toward practical privacy-preserving analytics
  for {IoT} and cloud-based healthcare systems. IEEE Internet Computing
  \textbf{22}(2),  42--51 (2018). \doi{10.1109/MIC.2018.112102519}

\bibitem{song.etal_2000_practical}
Song, D.X., Wagner, D., Perrig, A.: Practical techniques for searches on
  encrypted data. In: Proceeding 2000 {IEEE} symposium on security and privacy.
  pp. 44--55. IEEE, Berkeley, USA (2000). \doi{10.1109/SECPRI.2000.848445}

\bibitem{spiekermann.novotny_2015_vision}
Spiekermann, S., Novotny, A.: A vision for global privacy bridges: {Technical}
  and legal measures for international data markets. Computer Law \& Security
  Review: The International Journal of Technology Law and Practice
  \textbf{31}(2),  181--200 (2015)

\bibitem{sweeney_2002_kanonymity}
Sweeney, L.: k-anonymity: A model for protecting privacy. International Journal
  of Uncertainty, Fuzziness and Knowledge-Based Systems  \textbf{10}(05),
  557--570 (2002)

\bibitem{syverson.etal_1997_anonymous}
Syverson, P., Goldschlag, D., Reed, M.: Anonymous connections and onion
  routing. In: Proceedings. 1997 {IEEE} symposium on security and privacy. pp.
  44--54 (1997)

\bibitem{thurmond_2001_point}
Thurmond, V.A.: The point of triangulation. Journal of Nursing Scholarship
  \textbf{33}(3),  253--258 (2001). \doi{10.1111/j.1547-5069.2001.00253.x}

\bibitem{trask.etal_2020_privacy}
Trask, A., Bluemke, E., Garfinkel, B., Ghezzou Cuervas-Mons, C., Dafoe, A.:
  Beyond privacy trade-offs with structured transparency (2020)

\bibitem{vombrocke.etal_2009_reconstructing}
Vom~Brocke, J., Simons, A., Niehaves, B., Riemer, K., Plattfaut, R., Cleven,
  A.: Reconstructing the {Giant}: {On} the {Importance} of {Rigour} in
  {Documenting} the {Literature} {Search} {Process}. In: Proceedings of the
  17th {European} {Conference} on {Information} {Systems}. pp. 1--12. AIS,
  Verona, Italy (2009)

\bibitem{webster.watson_2002_analyzing}
Webster, J., Watson, R.T.: Analyzing the {Past} to {Prepare} for the {Future}:
  {Writing} a {Literature} {Review}. MIS Quarterly  \textbf{26}(2),  xiii --
  xxiii (2002)

\bibitem{yao_1982_protocols}
Yao, A.C.: Protocols for secure computations. In: 23rd annual symposium on
  foundations of computer science. pp. 160--164. IEEE, Chicago, IL, USA (1982)

\bibitem{zrenner.etal_2019_usage}
Zrenner, J., Möller, F.O., Jung, C., Eitel, A., Otto, B.: Usage control
  architecture options for data sovereignty in business ecosystems. Journal of
  Enterprise Information Management  \textbf{3}(32),  477--495 (2019)

\bibitem{zoll.etal_2021_privacysensitive}
Zöll, A., Olt, C.M., Buxmann, P.: Privacy-sensitive business models:
  {Barriers} of organizational adoption of privacy-enhancing technologies. In:
  Proceedings of the 29th european conference on information systems. pp.
  1--21. AIS (2021)

\end{thebibliography}

\end{document}